\documentstyle[11pt,paspconf,epsfig]{article}

\begin{document}

\title{The new Local Group galaxy candidates Cas dSph, Peg dSph=AndVI, and Cam~A}
\author{C. Gallart\altaffilmark{1}, D. Mart\'\i nez-Delgado\altaffilmark{2}, A. Aparicio\altaffilmark{2} and W.L. Freedman\altaffilmark{1}}
\affil{1. Carnegie Observatories. 813 Santa Barbara St. Pasadena CA 91101, USA}
\affil{2. Instituto de Astrof\'\i sica de Canarias. E-38200 La Laguna, Canary Islands, Spain}

\begin{abstract}
We present observations of the new Local Group galaxy candidates Cassiopeia dSph, Pegasus dSph = And VI and Camelopardalis A. Our deep color-magnitude diagrams show that the first two galaxies are certainly Local Group members, and likely dSph galaxies at a distance similar to that of the Andromeda galaxy. Cam A seems to be a star forming galaxy situated considerably further away.
\end{abstract}


\section{Introduction}

In the August 1998 issue of the {\it Dwarf Tales} Newsletter, Karatchensev and Karatchenseva reported the discovery of two new probable companions to M~31 in the Pegasus and Cassiopeia constellations, which were identified in their visual inspection of the POSS-II film copies. One of these galaxies had been discovered independently by Armandroff et al. (as reported in the same issue) using a digital filtering technique on POSS-II images. They provisionally designated their galaxy And VI. In the above mentioned issue of {\it Dwarf Tales} it was not clear which one of the two galaxies reported by Karachentsev and Karachentseva corresponded to And VI. In our case, this was only to be discovered right at the telescope! At the London airport on our way to South Africa, the excitement produced by the recent discovery of these new Local Group galaxies was clear when nobody could wait to attend the meeting to share all the new images and color-magnitude diagrams (CMDs).

After more work and some thought, we present here our data on these new Local Group galaxies, together with the CMD of another Local Group galaxy candidate, Cam A, discovered by Karatchensev (1994).     

\section{Observations and data reduction.}

Peg dSph=And VI, Cas dSph, and Cam A were observed at the Palomar 200" telescope (Palomar Observatory, USA) on August 25 and 26, 1998. The seeing was excellent, 0.8\arcsec--0.9\arcsec. Only the first night, when Peg dSph and Cam A were observed, was photometric. A photometric transformation for Cas dSph was obtained later at the 80 cm telescope IAC80 at the Teide Observatory (Canary Islands, Spain). The photometry for all three galaxies was obtained using the DAOPHOT/ALLFRAME programs (Stetson 1994). For a complete description of the observations and data reduction, see Gallart et al. (1999b).

\section{Peg dSph=And VI and Cas dSph}

Peg dSph=And VI and Cas dSph are highly resolved into stars in our images. Their smooth, elliptical profile and their lack of blue stars and HII regions suggest that they may be two new dSph galaxies similar to the other four (known so far) Andromeda companions, AndI-III (Caldwell et al. 1992) and And V (Armandroff et al. 1998). Their CMDs, which show a well populated red-giant branch (RGB) but no signs of any recent star formation activity, confirm this.

\begin{figure}
\begin{center}
\mbox{\epsfig{file=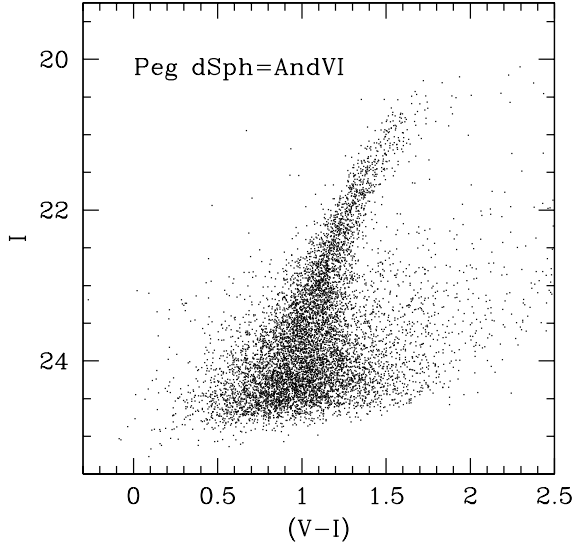,width=9.0cm,bbllx=1.5truecm,bblly=7.5truecm,bburx=21.0truecm,bbury=22.0truecm}}
\end{center}
\caption[]{$[(V-I), I]$ CMD of Peg dSph=AndVI.}
\label{peg}
\end{figure}

Figure~\ref{peg} displays the CMD of Peg dSph obtained with a total exposure time of 1 hour in both $V$ and $I$ with seeing under 0.8\arcsec. The tip of the RGB (TRGB) appears at $I\simeq 20.7\pm0.1$. Our photometry limit is at $I\simeq24.5$, i.e. about 4 magnitudes below the TRGB. Therefore, the horizontal-branch (HB) of an old, metal-poor population is expected to be near the limit of the photometry. In fact, the widening of the lower part of the CMD may be due not only to the observational errors, but also to the fact that we are reaching the HB and red-clump of core He-burning stars, which are more extended in color than the RGB.

\begin{figure}
\begin{center}
\mbox{\epsfig{file=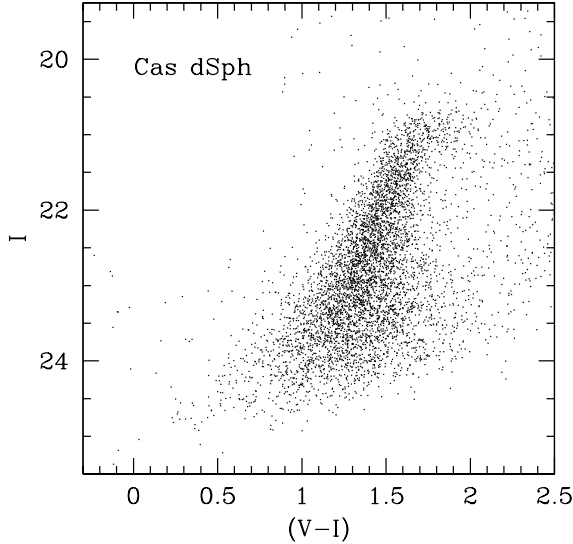,width=9.0cm,bbllx=1.5truecm,bblly=7.5truecm,bburx=21.0truecm,bbury=22.0truecm}}
\end{center}
\caption[]{$[(V-I), I]$ CMD of Cas dSph.}
\label{cas}
\end{figure}

Figure~\ref{cas} displays the CMD of the Cas dSph obtained with the same exposure time as the Peg dSph. Nevertheless, the fact that the night was not photometric and the seeing was slightly worse (but still as good as 0.9\arcsec) results in a limiting magnitude  $I\simeq 24$.
The TRGB is at $I=20.7\pm0.1$.

From the magnitude of the TRGB, a distance similar to the distance of the Andromeda Galaxy (770 kpc, Freedman \& Madore 1990) can be inferred for both Peg dSph and Cas dSph. This, together with the position of these galaxies in the sky, may point to a physical connection with M31. 

The CMD of both galaxies indicates a predominantly old stellar population with some intermediate-age population suggested by the presence of a few stars above
the TRGB. In the case of the Peg dSph the position and width of the RGB seem to indicate a very low metallicity; since in that case, the AGB would be less extended, we cannot exclude the possibility of a relatively large intermediate-age population fraction. This is the case for the predominantly intermediate-age dSph galaxy Leo I (Gallart et al. 1999a), whose TRGB region is quite similar to the one presented by Peg dSph. Cas dSph appears to have a considerably larger metallicity and metallicity dispersion than Peg dSph.   

\section{Cam A}

\begin{figure}
\begin{center}
\mbox{\epsfig{file=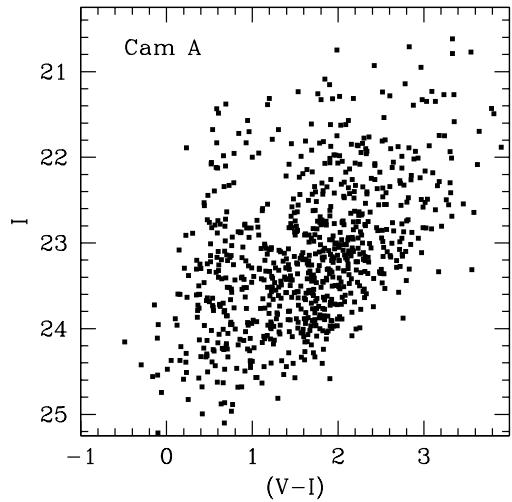,width=9.0cm,bbllx=1.5truecm,bblly=8truecm,bburx=21.0truecm,bbury=22.0truecm}}
\end{center}
\caption[]{$[(V-I), I]$ CMD of Cam A.}
\label{cama}
\end{figure}

Cam A appears much less resolved in our observations than the other two galaxies, and has an irregular morphology, with a large number of blue super-giant stars and a number of apparently semi-resolved knots which may be star clusters or HII regions. Its CMD in Figure~\ref{cama} is similar to the bright part of the CMD of star forming Local Group galaxies like M33, IC1613 or NGC 6822, with a blue plume of massive main-sequence stars at $0 \le (V-I) \le 1$ and a red-supergiant branch at $1.5 \le (V-I) \le 2.5$. The overdensity of stars fainter than $I=23$ at $(V-I)\simeq2$ may be the red-tangle (Gallart et al. 1996), observed near the limit of our photometry.

A crude estimate of the distance of this galaxy can be obtained from the different features in the CMD. First, from the magnitude of the brightest stars, assuming that they would have approximately the same $M_I$ as those in M33 and IC1613, we infer a distance modulus for Cam A of $(m-M) \simeq 28$. At this distance the TRGB would be detected at $I\simeq 24$ i.e., it would be at the limit of our photometry. Alternatively, if the overdensity of stars fainter than $I=23$ at $(V-I)\simeq2$  is indeed the red-tangle (Gallart et al. 1996), it would imply a TRGB at $I\simeq 23$, and a distance modulus of $(m-M) \simeq 27$.

The position of Cam A in the direction of the M81 group as defined by Karachentseva et al. (1985), and its distance as estimated here, are compatible with this galaxy being part of the M81 group.

\acknowledgments

C.G. and A.A acknowledge an IAU travel grant to attend this Symposium. Support for this work was provided by NASA grant GO-5350-03-93A and by the IAC (grant PB3-94).


\begin{references}

\reference Armandroff, T.E., Davies, J.E. \& Jacoby, G.H. 1998, \aj, 116, 2287

\reference Brinks, E. \& Grebel, E. (Eds) 1998, {\it Dwarf Tales}, 3 (http://www.astro.ugto.mx/$\sim$dwarfs) 

\reference Caldwell, N., Armandroff, T.E., Seitzer, P. \& Da Costa, G.S. 1992, \aj, 103, 840

\reference Freedman, W.L. \& Madore, B.F. 1990, \apj, 365, 186

\reference Gallart, C., Aparicio, A. \& V\'\i lchez, J.M. 1996, \aj, 112, 1928 

\reference Gallart, C., Freedman, W.L., Mateo, M., Chiosi, C., Thompson, I.B., Aparicio, A., Bertelli, G., Hodge, P., Lee, M.Y., Olsewszki, E.O., Saha, A., Stetson, P.B. \&  Suntzeff, N. 1999a, \apj, 514 (2) 

\reference Gallart, C.  et al. 1999b, in prep.

\reference Karachentsev, I.D. 1994, Astron. \& Astrophys. Transac. 6, 1

\reference Karachentseva, V.E. Karachentsev, I.D. \& B\"orngen, F. 1985, \aaps, 60, 213


\reference Stetson P. B. 1994, \pasp, 106, 250

\end{references}
\end{document}